\documentclass[pra,twocolumn,amssymb]{revtex4}
\usepackage{graphicx}
\usepackage{ulem}

\begin{document}

\title{Stability of a Fully Magnetized Ferromagnetic state in Repulsively Interacting Ultracold Fermi Gases}
\author{Xiaoling Cui$^{1}$ and Hui Zhai$^{2}$}
\affiliation{$^1$ Beijing National Laboratory for Condensed Matter Physics and Institute for Physics, Chinese Academy of Science, Beijing, 100190, China \\
$^2$ Institute for Advanced Study, Tsinghua University, Beijing, 100084, China}
\date{\today}

\begin{abstract}

We construct a variational wave function to study whether a fully
polarized Fermi sea of ultracold atoms is energetically stable
against a single spin flip. Our variational wavefunction contains
short-range correlations at least to the same level as
Gutzwiller's projected wavefunction. For the Hubbard lattice model and
the continuum model with pure repulsive interaction, we show a fully
polarized Fermi sea is generally unstable even for infinite
repulsive strength. By contrast for a resonance model, the ferromagnetic state
is possible if the $s$-wave scattering length is positive and
sufficiently large and the system is prepared to be orthogonal to
molecular bound state. However, we can not rule out the possibility
that more exotic correlation can destabilize the ferromagnetic
state.
\end{abstract}

\maketitle

Whether a fermion system with repulsive interaction will become
ferromagnetic is a long-standing problem in condensed matter
physics. Early in 1930s', Stoner used a simple mean-field theory to
predict that ferromagnetism will always take place with sufficient
large repulsive interaction \cite{Stoner}. However, this conclusion
is later challenged by Gutzwiller who took the short-range
correlation into account \cite{Gutzwiller}. So far, except a few specific cases \cite{Nagaoka,specific}, there is no
conclusive result on itinerant ferromagnetism. Recently, MIT group
reports an experiment on itinerant fermions in an ultracold Fermi
gas with large positive scattering length close to a Feshbach
resonance \cite{Ketterle}, and they attribute their observations to
Stoner ferromagnetism by comparing to theories
\cite{theory,MacDonald}. However, these theories are basically
mean-field theory or a second-order perturbation, which neither
include the Gutzwiller type short-range correlation nor consider the
unitary limited interaction nearby Feshbach resonance. Moreover many of the experimental signatures can be reproduced qualitatively by a non-magnetic correlated state \cite{Zhai}. Thus, it
calls for a serious study including the effects of both correlation
and unitarity in this problem.

In this Rapid Communication we address the question whether a fully magnetized
state is stable against a single spin flip. We compare the energy of
$N+1$ spin-up particles with that of one spin-down particle and $N$
spin-up particles. A fully magnetized ferromagnetic state is
definitely unstable if we can find a variational state of the latter
whose energy is lower. Similar idea has been used previously in
studying the stability of Nagaoka ferromagnetism in the Hubbard
model \cite{Anderson,Basile,Roth,Nagaoka}, and attractively
interacting Fermi gases with large population imbalance
\cite{Chevy,imbalance}. In this work, we will explore different
realizations of ``repulsive interactions" in ultracold Fermi gases:

({\bf I}) Single-band Hubbard model in a two-dimensional (2D) square
or three-dimensional (3D) cubic lattice. The Hamiltonian
$\hat{H}=\hat{H}_{\text{t}}+\hat{H}_{\text{int}}$,
$\hat{H}_{\text{t}}=-t\sum_{\langle
ij\rangle,\sigma}c^\dag_{i\sigma} c_{j\sigma}+\text{h.c}.$, where $\langle
ij\rangle$ are nearest-neighbor sites, and
$H_{\text{int}}=U\sum_{i}n_{i\uparrow}n_{i\downarrow}$ ($U>0$).

({\bf II}) Continuum model with finite-range interaction
potential in three dimension. $\hat{H}=\sum_{{\bf k}\sigma}\epsilon_{{\bf
k}}c^\dag_{{\bf k}\sigma}c_{{\bf k}\sigma}+(1/\Omega)\sum_{{\bf
q,k,k^\prime}}V({\bf k-k^\prime})c^\dag_{{\bf
q+k}\uparrow}c^\dag_{{\bf q-k}\downarrow}c_{{\bf
q-k^\prime}\downarrow}c_{{\bf q+k^\prime}\uparrow}$. Here
$\epsilon_{{\bf k}}= {\bf k}^2/(2m)$, $\Omega$ is system volume, and
$V({\bf k})$ is the Fourier transformation of real space interaction
potential $V({\bf r})$. Let $r_0$ be the interaction range, and thus
$V({\bf k})\rightarrow 0$ for $|{\bf k}|\gg k_0=1/r_0$. For the
convenience of later calculations, we adopt $s$-wave separable
potential $V({\bf k-k^\prime})=Uw({\bf k})w({\bf k^\prime})$, and
approximate $w({\bf k})=1/\sqrt{1+e^{\alpha(|{\bf k}|-k_0)/k_0}}$
with $\alpha\gg 1$. The $s$-wave scattering length $a_{\text{s}}$ is
related to $U$ as $m/(4\pi a_{\text{s}})=1/U+(1/\Omega)\sum_{|{\bf
k}|=0}^{k_c}1/(2\epsilon_{{\bf k}})$, where $k_c=k_0 \ln
(1+e^\alpha)/\alpha$. For a repulsive interaction $U>0$, $a_s$ is positive but upper bounded by $\pi/(2k_c)$ at $U\rightarrow +\infty$, and no bound state exists; for an attractive
$U<0$, $a_{\text{s}}$ diverges at
$U_c=-2\pi^2/(mk_c)$, and only a sufficient attraction $U<U_c$ with $a_{\text{s}}>0$ can support a two-body bound state.

({\bf III}) Continuum model with zero-range interaction
potential in three dimension. $\hat{H}=\sum_{{\bf k}}\epsilon_{{\bf k}}c^\dag_{{\bf
k}\sigma}c_{{\bf k}\sigma}+g\sum_{{\bf q,k,k^\prime}}c^\dag_{{\bf
q+k}\uparrow}c^\dag_{{\bf q-k}\downarrow}c_{{\bf
q-k^\prime}\downarrow}c_{{\bf q+k^\prime}\uparrow}$, and $g$ is
related to $a_{\text{s}}$ by $m/(4\pi
a_{\text{s}})=1/g+(1/\Omega)\sum_{|{\bf
k}|=0}^{\infty}1/(2\epsilon_{{\bf k}})$. There is always a two-body bound
state when $a_{\text{s}}>0$.

The single-band Hubbard model, as a simplified model for cold atoms in optical lattices (valid when interaction smaller than band gap) and many correlated materials, has been extensively studied before\cite{Anderson, Basile, Roth, Nagaoka}.
The comparison to previous known results justifies the validity of our
approach and calibrates the correlation incorporated in our
variational wavefunction (w.f.). Then we apply our method to continuum models which are
commonly-used for quantum gases and are of our primary interests.

The variational w.f.  we used for one down-spin
system  is similar to that used in the discussion of imbalanced
Fermi gases \cite{Chevy}, which is
\begin{equation}
|\Psi\rangle=\left(\phi_0 c^\dag_{{\bf
q_0}\downarrow}+\sum\limits_{{\bf k>k_{\text{F}}},{\bf
q<k_{\text{F}}}}\phi_{{\bf k}{\bf q}}c^\dag_{{\bf q_0+
q-k}\uparrow}c^\dag_{{\bf k}\uparrow}c_{{\bf
q}\uparrow}+\dots\right)|N\rangle\label{wf}
\end{equation}
where $|N\rangle$ represents a Fermi sea of $N$-spin up particles
with Fermi momentum $k_{\text{F}}$. ``$\dots$" in (\ref{wf})
represents terms contains more than one particle-hole pairs of
spin-up particles. We compute the energy $\mathcal{E}$ (measured
from energy of $|N\rangle$) of $|\Psi\rangle$, and compare it with
$E_{\text{F}}$ ($=E_{|N+1\rangle}-E_{|N\rangle}$). Our main results are summarized as follows.

First, for the single-band Hubbard model, we show that under certain
conditions, $H_{\text{int}}|\Psi\rangle=0$ and $\mathcal{E}=\langle
\Psi|H_{\text{t}}|\Psi\rangle<E_{\text{F}}$ for most range of
particle filling, except for nearby half-filling where the ground state is
rigorously proved to be ferromagnetic by Nakaoka \cite{Nagaoka}.
(See Fig. 1). This result agrees with previous studies by various
other methods \cite{Anderson,Basile,Roth}; We show that the real
space representation of $|\Psi\rangle$ corresponds to Gutzwiller's
w.f. with optimized ``backflow" type corrections, which implies $|\Psi\rangle$ includes short-range correlations at least as the Gutzwiller projection. Similarly for the finite-range continuum model, we find $\mathcal{E}<E_{\text{F}}$ for all range of $U>0$.
(See Fig. 2). This shows that generally in a purely repulsive
interaction model, a fully magnetized ferromagnetic state can not be
ground state even for infinite $U$, which is in sharp contrast to Stoner's mean-field conclusion.

Second, for the zero-range continuum model, apart from the polaron state with
negative energy discussed before in Ref.
\cite{Chevy,imbalance,MC,polaron}, we find the w.f. orthogonal to
the polaron state has a positive $\mathcal{E}$ for $a_{\text{s}}>0$.
At small $k_{\text{F}}a_{\text{s}}$, $\mathcal{E}$ as a function of
$k_{\text{F}}a_{\text{s}}$ follows the prediction of perturbation
expansion very well. At large $k_{\text{F}}a_{\text{s}}$,
$\mathcal{E}$ saturates (due to unitarity) to $1.82E_{\text{F}}$.
Below $k_{\text{F}}a^{\text{c}}_{\text{s}}=2.35$, we can find
variational state with which $\mathcal{E}<E_{\text{F}}$, but fail to
find such cases otherwise. (See Fig. 3). This result contradicts to
the previous prediction $k_{\text{F}}a^{\text{c}}_{\text{s}}=1.40$
based on perturbation theory \cite{MacDonald, notation}. Thus the
necessary condition for ferromagnetism is a large enough
$a_{\text{s}}\gtrsim 2.35/k_{\text{F}}$, which requires a
sufficiently attractive interaction potential to support a bound
state and cause resonant scattering, and more importantly the system
has to be prepared in the metastable scattering state that is
orthogonal to molecule state. Nevertheless, our method can not prove
this is a sufficient condition. Similar result is found for the finite-range model with $U<0$ and nearby scattering resonance. The details of our calculation are explained below.

{\it Single-band Hubbard model:} Here we can first prove

Theorem 1: Under two conditions that (i) $\phi_{{\bf k
q}}\equiv\phi_{\bf k}$ independent of ${\bf q}$, and (ii)
$\phi_0=-\sum_{{\bf k}>{\bf k_F}} \phi_{\bf k}$, the w.f.
\begin{equation}
|\psi\rangle_{1}=\left(\phi_0 c^\dag_{{\bf
q_0}\downarrow}+\sum_{{\bf k}>{\bf k_F}, {\bf q}<{\bf
k_F}}\phi_{{\bf k q}}c^\dag_{{\bf q_0}+{\bf q}-{\bf
k}\downarrow}c^\dag_{{\bf k}\uparrow}u_{{\bf q}\uparrow} \right)
|N\rangle,\label{general_psi}
\end{equation}
is an exact eigenstate of $\hat{H}_{\text{int}}$, and
$\hat{H}_{\text{int}}|\psi\rangle_{1}=0$.

This theorem can be verified straightforwardly. The subscript $_1$
of $|\psi\rangle$ means it contains one particle-hole pairs of
up-spins, and the best variational energy for this state is denoted
by $\mathcal{E}^{(1)}$. Condition (i) ensures that
$\hat{H}_{\text{int}}$ acting onto $|\psi\rangle_{1}$ will not
generate two particle-hole term, and condition (ii) ensures zero
interaction energy. For an intuitive understanding of this zero
interaction energy, we can Fourier transfer it to real space. Using
condition (ii), one can show the w.f. (\ref{general_psi}) is
equivalent to
\begin{eqnarray}
\left(\frac{\phi_0}{\sqrt{N_s}}\sum\limits_{{\bf m}}c^\dag_{{\bf
m}\downarrow}\mathcal{P}_{{\bf m}} +\sum\limits_{{\bf n}\neq {\bf
m}}\phi_{{\bf mn}}c^\dag_{{\bf m}\downarrow}c^\dag_{{\bf n}\uparrow}
c_{{\bf m}\uparrow}\right)e^{i{\bf q_0}{\bf m}}|N\rangle\nonumber.
\end{eqnarray}
Here $\mathcal{P}_{{\bf m}}=1-c^\dag_{{\bf m}\uparrow}c_{{\bf
m}\uparrow}$ presents standard Gutzwiller projection operator. $N_{\text{s}}$ is the number of lattice sites.
$\phi_{{\bf mn}}=\sum_{{\bf k}>{\bf k_F}}\phi_{{\bf k}}e^{i{\bf
k}{\bf (m-n)}}$, and the second term presents ``backflow" type
corrections. It becomes obvious that there will be no double
occupancy and no interaction energy. Hence, we have established a
momentum space w.f. representation of short-range correlation, which
can be generalized to free space straightforwardly.

\begin{figure}[btp]
\includegraphics[height=3.5cm,width=5.5cm]{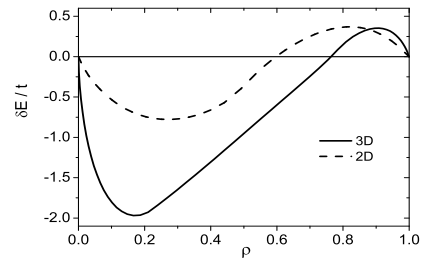}
\caption{$\delta E=\mathcal{E}^{(1)}-E_{\text{F}}$(in units of $t$)
as a function of particle density $\rho=N/N_{\text{s}}$ in a 3D cubic (solid) and a
2D square (dashed line) lattice. \label{Hubbard}}
\end{figure}

Next we try to find the minimum total(=kinetic) energy by
introducing a Lagrange multiplier that can be proved as just the
energy $\mathcal{E}^{(1)}$ (counted from the band bottom),
$\mathcal{F}=\langle\psi|\hat{H}_t|\psi\rangle-\mathcal{E}^{(1)}
\langle\psi|\psi\rangle$. Minimization of $\mathcal{F}$ gives ${\bf
q_0}=0$ and the self-consistent equation
\begin{equation}
\sum\limits_{{\bf
k>k_{\text{F}}}}\frac{\mathcal{E}^{(1)}}{\sum_{{\bf q<k_{\text{F}}}}
(\epsilon_{{\bf k q}}-\mathcal{E}^{(1)})}=1,
\end{equation}
in which $\epsilon_{{\bf k q}}=\epsilon_{{\bf q-k}}+\epsilon_{{\bf
k}}-\epsilon_{{\bf q}}$. Its solution gives $\mathcal{E}^{(1)}$.

$\delta E=\mathcal{E}^{(1)}-E_{\text{F}}$ as a function of particle
density $\rho$ in a cubic and a square lattice is plotted in Fig.
\ref{Hubbard}. We find for $\rho<\rho_c$, $\delta E<0$ means the
ferromagnetic state is always unstable even for infinite $U$. In fact, for a cubic
lattice, when $\rho\rightarrow 0$, $\mathcal{E}^{(1)}$ goes like
$\rho$ while $E_{\text{F}}$ goes like $\rho^{2/3}$, hence $\delta E$
is negative and shows a much rapid decrease compared with the square
lattice. While for $\rho>\rho_c$, $\delta E>0$, this is consistent
with Nagaoka theorem, which forces $\delta E$ to be positive when
$\rho\rightarrow 1$. $\rho_c$ obtained as $0.59\ (0.76)$ for the
square (cubic) lattice agree with previous studies by a finite-size
real space evaluation\cite{Basile} or by using Green
functions\cite{Roth}. By contrast, our variational w.f. (Eq.
\ref{general_psi}) greatly simplifies the calculation, and also
enable calculations in the thermodynamic limit simply by employing
density of state in a one-dimensional integral equation. Moreover,
this w.f. can be systematically improved by including multiple
particle-hole contributions.

Theorem 2: Consider the wave function $|\psi\rangle_{n}$ that
contains up to $n$ particle-hole pairs
\begin{eqnarray}
\sum\limits_{m=0}^{n}\frac{1}{(m!)^2}\sum_{\{ {\bf k_i} \}_{1}^{m};
\{ {\bf q_j}\}_{1}^{m}}\phi^{(m)}_{\{ {\bf k_i}\}_{1}^{m};\{ {\bf
q_j}\}_{1}^{m}}c^\dag_{{\bf
p_m}\downarrow}\prod_{i=1}^{m}c^\dag_{{\bf
k_i}\uparrow}\prod\limits_{j=1}^{m}c_{{\bf q_j}\uparrow}
|N\rangle\nonumber
\end{eqnarray}
with ${\bf p_m}={\bf q_0}+\sum_{i=1}^m({\bf q_i}-{\bf k_i})$, where
$\{ {\bf k_i} \}_{1}^{m}$ denotes a set $\{ {\bf k_1},\dots,{\bf
k_m}\}$ and $\{ {\bf q_j} \}_{1}^{m}$ denotes $\{ {\bf
q_1},\dots,{\bf q_m}\}$. It satisfies
$H_{\text{int}}|\psi\rangle_{n}=0$ under the condition (i)
$\phi^{(m)}_{\{ {\bf k_i}\}_{1}^{m};\{ {\bf q_j} \}_{1}^{m}}$ can be
expressed as $B^{(m)}_{\{ {\bf k_i} \}_{1}^{m};{\bf
q_{m-1}},\dots,{\bf q_1}}-B^{(m)}_{\{ {\bf k_i} \}_{1}^{m};{\bf
q_{m}},{\bf q_{m-2}},\dots,{\bf q_1}}+\dots+(-1)^{m-1}B^{(m)}_{\{
{\bf k_i} \}_{1}^{m};{\bf q_{m}},{\bf q_{m-1}},\dots,{\bf q_2}}
+C^{(m)}_{\{ {\bf k_i} \}_{1}^{m};\{ {\bf q_j}\}_{1}^{m}}, $ with
$B^{(0)}=0$, $C^{(0)}=\phi^{(0)}$; $C^{(n)}_{\{ {\bf
k_i}\}_{1}^{m};\{ {\bf q_j}\}_{1}^{m}}=0$; and (ii) $C^{(m-1)}_{\{
{\bf k_i} \}_{1}^{m-1};\{ {\bf q_j} \}_1^{m-1}}+\sum_{{\bf
k_{m}}}B^{(m)}_{\{{\bf k_i} \}_{1}^{m-1},{\bf k_m}; \{ {\bf
q_j}\}_{1}^{m-1}}=0$ for any $\{ {\bf k_i} \}_{1}^{m-1}$ and $\{
{\bf q_j}\}_{1}^{m-1}$.

Similar to Theorem 1, this theorem can be verified
straightforwardly, and condition (i) ensures $H_{\text{int}}$ acting
on $|\psi\rangle_n$ will not generate $n+1$ particle-hole terms, and
condition (ii) ensures zero-energy. Obviously, $|\psi\rangle_{n-1}$
is a special case of $|\psi\rangle_n$, and hence
$\mathcal{E}^{(n)}\leq \mathcal{E}^{(n-1)}$, where
$\mathcal{E}^{(n)}$ is the best variational energy for
$|\psi\rangle_n$. By including more particle-hole terms, one can
always further lower $\mathcal{E}$, and make $\rho_c$ more close to
unity. On the other hand, one can also prove $\lim_{\rho\rightarrow
0}(\mathcal{E}^{(n)}-\mathcal{E}^{(n-1)})/\mathcal{E}^{(1)}
\rightarrow 0$, namely, the multiple particle-hole contribution
vanishes and $\mathcal{E}^{(1)}$ becomes exact at low density limit.

\begin{figure}[btp]
\includegraphics[height=4.5cm,width=8.2cm]{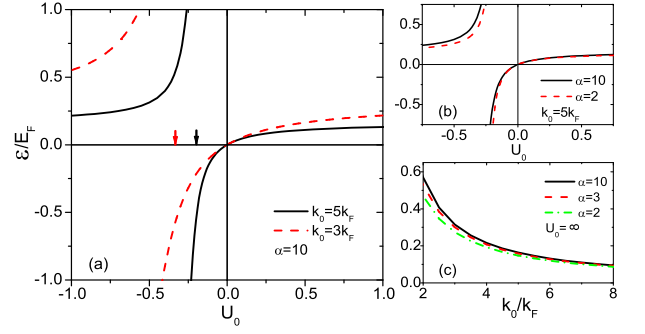}
\caption{(color online). $\mathcal{E}/E_{\text{F}}$
for the continuum model with finite-range interactions. (a)
$\mathcal{E}$ as a function of $U_0$ for $k_{\text{0}}/k_{\text{F}}=2$ or $5$,
$\alpha=10$. Arrows denote the corresponding critical $U_c$ that can
afford a two-body bound state. (b) the same plot as (a) for
$k_{\text{0}}/k_{\text{F}}=5$ and $\alpha=2$ or $10$. (c) $\mathcal{E}/E_{\text{F}}$ at
$U_0\rightarrow +\infty$ as a functions $k_0$ for $\alpha=2,3,10$.
\label{finiterange}}
\end{figure}

{\it Finite-range continuum model:} We consider
the same variational w.f. as Eq. \ref{general_psi} with $\phi_{{\bf
k}{\bf q}}\equiv\phi_{\bf k}$. The saddle point equation from energy
minimization is
\begin{eqnarray}
\frac{U}{\Omega}\sum_{\bf q}\left( \phi_0 +\sum_{{\bf k}}w_{\bf k}\phi_{{\bf k}}\right)&=&\mathcal{E}\phi_0,\nonumber\\
\frac{U}{\Omega}w_{{\bf k}}\sum_{{\bf
q}}\left(\phi_0+\sum\limits_{{\bf k^\prime}}w_{{\bf
k^\prime}}\phi_{{\bf k^\prime}}\right)&=&\sum_{{\bf
q}}(\mathcal{E}-\epsilon_{{\bf k q}}) \phi_{{\bf k}},\label{sadd}
\end{eqnarray}
where we have adopted the approximation $w_{{\bf q}}\approx 1$ for
$|{\bf q}|<|{\bf k_{\text{F}}}|<k_0$. From Eqs. \ref{sadd} one can
obtain a self-consistent equation
\begin{equation}
\frac{2E_{\text{F}}}{3\mathcal{E}}-\frac{1}{U_0}=
\int_{1}^{+\infty}\frac{\tilde{k}^2 w^2(\tilde{k})
d\tilde{k}}{\tilde{k}^2-\mathcal{E}/(2E_{\text{F}})},\label{finite}
\end{equation}
where $\tilde{k}=k/k_{\text{F}}$ and $U_0=U mk_{\text{F}}/(2\pi^2)$
are dimensionless. The solution to Eq. \ref{finite} is plotted in
Fig. \ref{finiterange}(a). Fig. \ref{finiterange}(b) shows the
results are insensitive to the choice of the free parameter $\alpha$
in the separable potential.  As one can see, for $U_0\rightarrow
+\infty$, $\mathcal{E}$ saturates to a finite value, which is
smaller than $E_{\text{F}}$. This saturation precisely reflects the
physics that the repulsive interaction can be strongly renormalized
by short-range correlations, whose upper bound is the kinetic energy
cost for the screening, and is always finite even when the bare
interaction diverges. In fact, for $k_0\gg k_{\text{F}}$, the
integral at r.h.s. of Eq. \ref{finite} is approximately
$k_0/k_{\text{F}}$, hence at $U\rightarrow +\infty$,
$\mathcal{E}/E_{\text{F}}= 2k_{\text{F}}/(3k_0)<1$. For a general
case, we show in Fig. \ref{finiterange}(c)
$\mathcal{E}/E_{\text{F}}$ at $U_0=+\infty$ as a function of
$k_0/k_{\text{F}}$ at different $\alpha$, and this ratio is always
below unity.

\begin{figure}[tp]
\includegraphics[height=4.2cm,width=6.3cm]{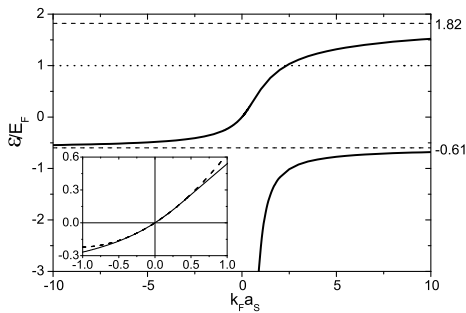}
\caption{$\mathcal{E}/E_{\text{F}}$ as a function of
$k_{\text{F}}a_{\text{s}}$ for the continuum model with zero range
interactions. Dashed horizontal lines denote saturated values of
$\mathcal{E}/E_{\text{F}}$ at unitarity. Inset is fit to the
perturbation results (see text) for small
$k_{\text{F}}a_{\text{s}}$. \label{zero-range}}
\end{figure}

{\it Zero-range continuum model:} This model is in fact the $k_0\rightarrow
+\infty$ limit of the finite-range model with $U<0$. 
For the w.f. of Eq.
\ref{general_psi}, $H|\psi\rangle=\mathcal{E}|\psi\rangle$ gives the
same self-consistent equation as in Ref. \cite{Chevy},
\begin{equation}
\frac{\mathcal{E}}{E_{\text{F}}}=2\int_0^1d\tilde{q}\tilde{q}^2\left[\frac{\pi}{2}\frac{1}{k_{\text{F}}a_{\text{s}}}-1+g\left(\frac{\mathcal{E}}{E_{\text{F}}},\tilde{q}\right)\right]^{-1}
\end{equation}
where $g(\mathcal{E}/E_{\text{F}},\tilde{q})$ is given by the integral
\begin{eqnarray}
\int_1^{+\infty}d\tilde{k}\left(\int_0^{\pi}d\theta\frac{\tilde{k}^2
\sin\theta}{2\tilde{k}^2-2\tilde{q}\tilde{k}\cos\theta-\mathcal{E}/E_{\text{F}}}-1\right)\nonumber
\end{eqnarray}
where $\tilde{q}=q/k_{\text{F}}$ and $\tilde{k}=k/k_{\text{F}}$. For this model there is always a bound state in a two-body problem for $a_{\text{s}}>0$, and correspondingly, there is always a polaron
solution with negative $\mathcal{E}$ when $a_{\text{s}}>0$, which
has been extensively discussed in Ref. \cite{Chevy,imbalance, MC,
polaron}. In addition, there is always a positive energy solution
orthogonal to the polaron solution. For small $k_Fa_s$ applying the second-order perturbation theory\cite{LHY} to this case, one will obtain
$\mathcal{E}/E_{\text{F}}=4k_{\text{F}}a_{\text{s}}/(3\pi)+2(k_{\text{F}}a_{\text{s}})^2/\pi^2$.
It is found our variational results fit quite well to this expansion
in the regime $|k_{\text{F}}a_{\text{s}}|\lesssim 0.6$ (see inset of
Fig. \ref{zero-range}), and start to deviate substantially when
$k_{\text{F}}a_{\text{s}}\gtrsim0.6$. At
$k_{\text{F}}a_{\text{s}}\rightarrow+\infty$, we find the energy
under this trial w.f. saturates at $1.82E_{\text{F}}$ instead of
diverging as perturbation theory predicts. This saturation is due to
unitary limit of resonance interaction. We find
$\mathcal{E}<E_{\text{F}}$ for $k_{\text{F}}a_{\text{s}}<2.35$,
which means a fully polarized ferromagnetism is definitely unstable
below a critical $k_{\text{F}}a^{\text{c}}_{\text{s}}=2.35$. It
contradicts to previous results based on a second-order perturbation
\cite{MacDonald} which predicts the system becomes fully magnetized
at $k_{\text{F}}a_{\text{s}}=1.40$ \cite{notation}. The reason for
this discrepancy is because the second order perturbation
overestimates the interaction effects in the regime
$k_{\text{F}}a_{\text{s}}\gtrsim 1$. Similar behavior is found for a
resonance scattering with finite interaction range, as shown in the
$U_0<0$ part of Fig. \ref{finiterange}(a).

We would like to point out several intrinsic relations between (a)
single-band Hubbard model at $U\gg t$, (b) finite-range continuum model at $U_0=+\infty$
and (c) zero-range continuum model at $a_{\text{s}}=+\infty$. First is the
relation between the coefficient of different terms in the w. f. (Eq.
{\ref{general_psi}}): for (b) we have $\phi_0+\sum_{{\bf
k>k_{\text{F}}}}w_{{\bf k}}\phi_{{\bf k}}=0$; this is equivalent to
(a) with all $w_{{\bf k}}=1$ and a momentum cut-off imposed by
$\pi/a_{\text{L}}$ instead of $k_0$ ($a_{\text{L}}$ is the lattice
constant); for (c) we can define $\chi_{{\bf q}}=\phi_0+\sum_{{\bf
k>k_{\text{F}}}}\phi_{{\bf k}{\bf q}}$ then we have $\phi_{{\bf
k}{\bf q}}=\chi^0_{{\bf q}}/(\mathcal{E}-\epsilon_{{\bf k}{\bf q}})$
with $\chi^0_{{\bf q}}=g\chi_{{\bf q}}/\Omega$\cite{tan}, so at
unitary limit one can get similar relation as $\phi_0+\sum_{{\bf
k>k_{\text{F}}}}(\phi_{{\bf k}{\bf q}}+\chi^0_{{\bf
q}}/(2\epsilon_{{\bf k}}))+\sum_{{\bf
q<k_{\text{F}}}}1/(2\epsilon_{{\bf q}})=0$. Second,
for (a) at low density limit and (b) with $k_F\ll k_0$ we can show
$\mathcal{E}/E_F\sim k_Fr_{\text{eff}}$, here the effective range
$r_{\text{eff}}=a_{\text{L}}$ for (a), and $1/k_0$ for (b), which
directly leads to the instability of ferromagnetism in these limits;
while (c) does not fall into the class of pure repulsive interactions.

The calculation above uses the w.f. that contains only one
particle-hole pairs of up-spins. Including multiple particle-hole
pairs can systematically improve the results, lead to further lower
$\mathcal{E}$ and increased $k_{\text{F}}a^{\text{c}}_{\text{s}}$.
However, from the experience in studies of polaron branch, it is
found the single particle-hole w.f. can already produce a result
very close to Monte Carlo simulation \cite{MC} and later experiments
\cite{polaron}, and the reason for this perfect agreement is
understood as a nearly perfect destructive interference of
higher-order particle-hole contributions \cite{imbalance}. Hence it
is unlikely $\mathcal{E}/E_{\text{F}}$ at resonance can be reduced
from $1.82$ to below unity. Even though, we only prove ferromagnetic
is stable against single spin flip above a critical
$k_{\text{F}}a^{\text{c}}_{\text{s}}$, and we can not rule out the
possibility that a state with more down-spin can be energetically
more favorable due to more exotic correlations.

Our results bring forward two intriguing issues. (i) for $0.6\lesssim
k_{\text{F}}a_{\text{s}}<2.35$, the system can neither be fully
ferromagnetic nor be well described by perturbation
theory. It is in a very interesting strongly interacting quantum
phase with large ferromagnetic and/or short-range fluctuations; (ii)
for $k_{\text{F}}a_{\text{s}}>2.35$, our approach predicts the system will become
ferromagnetic. If it is true, what is the smoking gun experimental
evidence? If future experiments find it is not true, then it means the system contains much stronger correlations than discussed here.

{\it Acknowledgment}: We thank Z. Y. Weng, Y. P. Wang, S.
Chen, J. L. Song for valuable discussions. H.Z. is supported by the Basic
Research Young Scholars Program of Tsinghua University, NSFC Grant
No. 10944002 and 10847002. X.L.C. is supported by NSFC, CAS and
973-project of MOST.


\begin{thebibliography}{99}
\bibitem{notation}
$k_{\text{F}}$ in our paper is the Fermi momentum for a fully polarized state, thus differs by a factor of $2^{1/3}$ from that previously defined for each component in a non-magnetic state \cite{Ketterle, theory, MacDonald}.
Hence the prediction for fully polarized
state $k_{\text{F}}a_{\text{s}}=1.112$\cite{MacDonald} is
translated to $1.40$ to compare with our result (2.35). Our
result should be translated to 1.87 to compare with MIT experiment
\cite{Ketterle}.
\end{thebibliography}
\end{document}